\documentclass[aps,prd,onecolumn,groupedaddress,showpacs,nofootinbib,amssymb]{revtex4}
\usepackage{graphicx,bm,color}
\usepackage{amsmath}
\usepackage{amssymb}
\usepackage{amsfonts}

\newcommand{\be}{\begin{equation}}
\newcommand{\ee}{\end{equation}}
\newcommand{\bea}{\begin{eqnarray}}
\newcommand{\eea}{\end{eqnarray}}
\newcommand{\beaa}{\begin{eqnarray*}}
\newcommand{\eeaa}{\end{eqnarray*}}

\newcommand{\nn}{\nonumber \\}

\allowdisplaybreaks[4]

\begin{document}

\title{New massive spin two model on curved space-time}

\author{Satoshi Akagi$^1$, Yuichi Ohara$^1$ and Shin'ichi Nojiri$^{1,2}$}

\affiliation{
$^1$ Department of Physics, Nagoya University, Nagoya 464-8602, Japan \\
$^2$ Kobayashi-Maskawa Institute for the Origin of Particles and
the Universe, Nagoya University, Nagoya 464-8602, Japan 
}

\begin{abstract}
We have proposed a new ghost-free model with interactions of massive spin two particles 
in Phys.\ Rev.\ D {\bf 90} (2014) 043006 [arXiv:1402.5737 [hep-th]]. 
Although the model is ghost-free on the Minkowski space-time, it is not obvious whether 
or not this desirable property is preserved on curved space-time. 
In fact, Buchbinder et al. already pointed out that the Fierz-Pauli theory is not 
ghost-free on curved space-time without non-minimal coupling terms. 
In this paper, we construct a new theory of massive spin two particles with non-minimal 
coupling on curved space-time and show that the model can be ghost-free. 
Furthermore, we propose new non-minimal coupling terms.

\end{abstract}


\maketitle

\section{Introduction \label{Sec1}}

The theory of massive spin two particles has a long history. 
Fierz and Pauli studied the theories with arbitrary spin and
succeeded in formulating the model describing the free massive spin two particle in 
1939 \cite{Fierz:1939ix}. 
This theory is well known as the Fierz-Pauli theory.
Although the model is a free field theory, the construction is non-trivial because the 
mass term generally leads to a ghost and breaks the consistency as a quantum theory. 
Fierz and Pauli removed the ghost by tuning the relative values between the 
coefficients of the non-derivative quadratic terms. 
Since the Fierz-Pauli theory has already lost the gauge symmetry due to the mass 
term, it might be expected that arbitrary interactions could be allowed in the massive spin two 
field theory unlike in the massless theory. 
Boulware and Deser \cite{Boulware:1974sr}, however, suggested that non-linear terms 
generally generate another kind of ghost called the Boulware-Deser ghost. 
In addition to the ghost, another problem appears if we regard the massive spin two 
theory as an alternative theory of gravity. 
The prediction of the free massive spin two theory does not coincide with the free 
massless theory even in the massless limit. 
This fact was pointed out by van Dam, Veltman and Zakharov (vDVZ) 
\cite{vanDam:1970vg} although the discontinuity can be screened by some non-linear 
effects called the Vainshtein mechanism \cite{Vainshtein:1972sx}. 
(see, for example, Ref.~\cite{Luty:2003vm}).
After the work by Boulware and Deser, the studies of non-linear massive spin two 
theories had not progressed until 2002 because the appearance of  
the Boulware-Deser ghost suggested some kind of no-go theorem.

In 2002, Arkani-Hamed, Georgi, and Schwartz \cite{ArkaniHamed:2002sp} considered 
massive gravity as a low energy effective field theory and showed that some class of 
the infinite potential terms can make the cut-off scale larger. 
Eight years later, de Rham, Gabadadze, and Tolley succeeded in obtaining the formal 
expression of the potential terms \cite{deRham:2010ik}
and it was proved that the potential-tuned theory is ghost-free 
\cite{Hassan:2011hr,Hassan:2011ea}.
This theory is called the dRGT massive gravity. 
The most essential part for the ghost-free property is the characteristic forms of the 
fully non-linear potential terms. 
As the dRGT massive gravity is accompanied with a non-dynamical metric called 
fiducial metric, we may consider to make the fiducial metric dynamical and obtain 
theories called bigravity which contain two dynamical metrics 
\cite{Hassan:2011zd,Hassan:2011vm,Hassan:2011tf}.

Recently Hinterbichler \cite{Hinterbichler:2013eza} suggested the possibility of new 
derivative interaction terms in the dRGT massive gravity by showing the existence of 
ghost-free derivative interactions for the Fierz-Pauli theory. 
It was shown that the leading term of the potential in the dRGT is also ghost-free as 
for the Fierz-Pauli theory. 
Based on this discussion, we constructed a new massive spin two model 
\cite{Ohara:2014vua} which contains a kinetic term and potential terms only. 

Although the new model is ghost-free on Minkowski space-time, it is not obvious whether or not 
the model on curved space-time keeps the property. 
The Fierz-Pauli theory coupled with gravity has been already discussed by 
Buchbinder et al. and they revealed that the minimal coupling model is not ghost-free 
\cite{Buchbinder:1999ar}. 
This is because some additional terms including the curvature tensor appear from 
non-commutativity of covariant derivatives and prevent the construction of 
the constraint. 
Therefore, non-minimal coupling terms are necessary so that the free massive spin two 
theory does not include any ghost when coupled with gravity. 
As a result, they formulated the ghost-free Fierz-Pauli theory on non-trivial 
background by adding two terms with non-minimal coupling and restricting the background to 
be the Einstein manifold.
This means our new model should contain at least two non-minimal couplings on curved 
space-time.  

In this paper, we consider the new massive spin two model coupled with gravity and add the 
two non-minimal coupling terms to the model. 
Then, we study whether or not the non-minimal coupling model is ghost-free using the Lagrangian 
formalism. Furthermore, we investigate another possibility of new ghost-free terms on the Einstein 
manifold and find a new class of ghost-free potential terms while it is shown to be impossible, in our 
formulation, to introduce derivative interactions without any ghost.

\section{New model of massive spin two particle}

In this section, we review on the model of the massive spin two particle with interaction 
proposed in \cite{Ohara:2014vua}. 
We start with the Lagrangian of the Fierz-Pauli theory \cite{Fierz:1939ix}: 
\be
\label{FPLag}
\mathcal{L}_{\mathrm{FP}} = -\frac{1}{2}\partial_\lambda h_{\mu\nu}\partial^\lambda 
h^{\mu\nu}+\partial_\mu h_{\nu\lambda}\partial^\nu h^{\mu\lambda}-\partial_\mu 
h^{\mu\nu}\partial_\nu h+\frac{1}{2}\partial_\lambda h\partial^\lambda h 
 -\frac{1}{2}m^2(h_{\mu \nu}h^{\mu \nu}-h^2) \, .
\ee
In order to eliminate the ghost, we need to tune the relative sign of the mass term. 
By Hinterbichler \cite{Hinterbichler:2013eza}, it has been pointed out that we may add 
new interaction terms to this model without generating any ghost by the specific linear 
combinations of the interaction terms. 
In four dimensions, we know that only two kinds of non-derivative interactions exist:
\begin{align}
\mathcal{L}_3 
\sim & \eta^{\mu_{1} \nu_{1} \mu_{2} \nu_{2} \mu_{3} \nu_{3}} h_{\mu_{1}\nu_{1}} h_{\mu_{2} 
\nu_{2}} h_{\mu_{3} \nu_{3}} \, ,
\label{nhhh} \\
\mathcal{L}_4
\sim & \eta^{\mu_{1} \nu_{1} \mu_{2} \nu_{2} \mu_{3} \nu_{3} \mu_{4} \nu_{4}} 
h_{\mu_{1} \nu_{1}} h_{\mu_{2} \nu_{2}} h_{\mu_{3} \nu_{3}} h_{\mu_{4} \nu_{4}} \, .
\label{nhhhh}
\end{align}
Here $\eta^{\mu_{1} \nu_{1} \cdots \mu_{n} \nu_{n}}$ is the product of $n$ 
$\eta_{\mu\nu}$ given by anti-symmetrizing the indices $\nu_1$, $\nu_2$, $\cdots$, and 
$\nu_n$ 
Some examples are given by, 
\begin{align}
\label{h3c}
\eta^{\mu_{1} \nu_{1} \mu_{2} \nu_{2}} \equiv &  
\eta^{\mu_{1} \nu_{1}} \eta^{\mu_{2} \nu_{2}} - \eta^{\mu_{1} \nu_{2}} \eta^{\mu_{2} 
\nu_{1}}\, , \nn
\eta^{\mu_{1} \nu_{1} \mu_{2} \nu_{2} \mu_{3} \nu_{3}} \equiv & 
\eta^{\mu_{1} \nu_{1}}\eta^{\mu_{2} \nu_{2}} \eta^{\mu_{3} \nu_{3}} - \eta^{\mu_{1} 
\nu_{1}}\eta^{\mu_{2} \nu_{3}} \eta^{\mu_{3} \nu_{2}}
+ \eta^{\mu_{1} \nu_{2}}\eta^{\mu_{2} \nu_{3}} \eta^{\mu_{3} \nu_{1}} \nn
& - \eta^{\mu_{1} \nu_{2}}\eta^{\mu_{2} \nu_{1}} \eta^{\mu_{3} \nu_{3}}
+ \eta^{\mu_{1} \nu_{3}}\eta^{\mu_{2} \nu_{1}} \eta^{\mu_{3} \nu_{2}} - \eta^{\mu_{1} 
\nu_{3}}\eta^{\mu_{2} \nu_{2}} \eta^{\mu_{3} \nu_{1}} 
\, .
\end{align}
The detailed property of $\eta^{\mu_1 \nu_1 \cdots \mu_n \nu_n}$ is summarized in Appendix A.
In \cite{Ohara:2014vua}, it was proposed a new model of massive spin two particles by 
adding the terms in (\ref{nhhh}) and (\ref{nhhhh}) to the Fierz-Pauli Lagrangian in 
(\ref{FPLag}), 
\begin{align}
\label{hh10}
\mathcal{L}_{h0} 
= & -\frac{1}{2} \eta^{\mu_{1} \nu_{1} \mu_{2} \nu_{2} \mu_{3} \nu_{3}} 
\left( \partial_{\mu_{1}} \partial_{\nu_{1}} h_{\mu_{2} \nu_{2}}\right) h_{\mu_{3} \nu_{3}}
 + \frac{m^2}{2} \eta^{\mu_{1} \nu_{1} \mu_{2} \nu_{2}} h_{\mu_{1} \nu_{1}} h_{\mu_{2} 
\nu_{2}} \nn
& - \frac{\mu}{3!} \eta^{\mu_{1} \nu_{1} \mu_{2} \nu_{2} \mu_{3} \nu_{3}} 
h_{\mu_{1}\nu_{1}} h_{\mu_{2} \nu_{2}} h_{\mu_{3} \nu_{3}}
 - \frac{\lambda}{4!} \eta^{\mu_{1} \nu_{1} \mu_{2} \nu_{2} \mu_{3} \nu_{3} \mu_{4} \nu_{4}} 
h_{\mu_{1} \nu_{1}} h_{\mu_{2} \nu_{2}} h_{\mu_{3} \nu_{3}} h_{\mu_{4} \nu_{4}} \nn
= & -\frac{1}{2} \left( h \Box h - h^{\mu\nu} \Box h_{\mu\nu} 
 - h \partial^\mu \partial^\nu h_{\mu\nu} - h_{\mu\nu} \partial^\mu \partial^\nu h 
+ 2 h_\nu^{\ \rho} \partial^\mu \partial^\nu h_{\mu\rho} \right) \nn
& + \frac{m^2}{2} \left( h^2 - h_{\mu\nu} h^{\mu\nu} \right) 
 - \frac{\mu}{3!} \left( h^3 - 3 h h_{\mu\nu} h^{\mu\nu} + 2 h_\mu^{\ \nu} h_\nu^{\ \rho} 
h_\rho^{\ \mu} \right)  \nn
& - \frac{\lambda}{4!} \left( h^4 - 6 h^2 h_{\mu\nu} h^{\mu\nu} + 8 h h_\mu^{\ \nu} 
h_\nu^{\ \rho} h_\rho^{\ \mu} 
 - 6 h_\mu^{\ \nu} h_\nu^{\ \rho} h_\rho^{\ \sigma} h_\sigma^{\ \mu} + 3 \left( 
h_{\mu\nu} h^{\mu\nu} \right)^2 \right) \, .
\end{align}
Here $m$ and $\mu$ are parameters with the dimension of mass although the parameter 
$\lambda$ is dimensionless. 
The parameter $\mu$ is always takes positive values but the sign of $\lambda$ is 
non trivial for the stabilities. 

We should note that the model with cubic interactions, including the derivatives 
interactions,  
was first proposed in \cite{Folkerts:2011ev} before \cite{Hinterbichler:2013eza} and it was 
also proved that there is no ghost in the model. 

\section{Pseudo-linear theory on flat space}

In this paper, we consider the model where the massive spin two field couples with gravity 
but  in order to construct the model without ghost on the curved space-time, 
we begin with the counting of the degrees of freedom on the flat space-time 
by using the lagrangian formalism as in \cite{Buchbinder:1999ar}.

First, just for simplicity, we only include the cubic interactions, that is, $\lambda=0$ in 
(\ref{hh10}). 
\begin{align}
\mathcal{L} = \frac12 \eta^{\mu_1 \nu_1 \mu_2 \nu_2 \mu_3 \nu_3 } \partial_{\mu_1} 
h_{\mu_2 \nu_2} \partial_{\nu_1} h_{\mu_3 \nu_3} 
+\frac{m^2}{2} \eta^{\mu_1 \nu_1 \mu_2 \nu_2} h_{\mu_1 \nu_1} h_{\mu_2 \nu_2} 
-\frac{\mu}{3!} \eta^{\mu_1 \nu_1 \mu_2 \nu_2 \mu_3 \nu_3} h_{\mu_1 \nu_1} 
h_{\mu_2 \nu_2 } h_{\mu_3 \nu_3}\, .
\label{A1}
\end{align}
By the variations with respect to $h_{\mu \nu} $, we obtain following equations,
\begin{align}
0=E_{\mu \nu} = -\eta_{( \mu \nu ) \mu_1 \nu_1 \mu_2 \nu_2} \partial^{\mu_1} 
\partial^{\nu_1} h^{\mu_2 \nu_2} + m^2 \eta_{\mu \nu  \mu_1 \nu_1}h^{\mu_1 \nu_1}
 -\frac{\mu}{2} \eta_{(\mu \nu ) \mu_1 \nu_1 \mu_2 \nu_2}  
h^{\mu_1 \nu_1} h^{\mu_2 \nu_2}\, . 
\label{e1}
\end{align}
In the equations (\ref{e1}), there are equations which contain the first order derivative 
with respect to time but do not contain the second order derivative, 
\begin{align}
0 = E_{0\nu} = -\eta_{(0 \nu ) \mu_1 \nu_1 \mu_2 \nu_2} \partial^{\mu_1} \partial^{\nu_1} 
h^{\mu_2 \nu_2} + m^2 \eta_{0 \nu  \mu_1 \nu_1}h^{\mu_1 \nu_1}
 -\frac{\mu}{2} \eta_{(0 \nu ) \mu_1 \nu_1 \mu_2 \nu_2}  
h^{\mu_1 \nu_1} h^{\mu_2 \nu_2} \, . 
\label{e2}
\end{align}
Because of the anti-symmetry of $\eta_{\mu_1 \nu_1 \mu_2 \nu_2 \mu_3 \nu_3}$, these 
equations do not contain any term including the second order derivatives with respect to 
time, ${\ddot{h}}_{\mu \nu}$ nor the first order derivatives of $h_{00}$ with respect to 
time.
Then we may regard $h_{0\mu}$ as auxiliary fields.
The remaining equations in (\ref{e1}) contains the second order derivative with respect to 
time, 
\begin{align}
0=E_{ij} = \eta_{(ij)kl} \ddot{h}_{kl} + (\text{terms without } \ddot{h})\, . 
\label{e3}
\end{align}
Here we used the following identity, 
\begin{align}
\eta_{\mu_1 \nu_1 \mu_2 \nu_2 \mu_3 \nu_3} = \eta_{\mu_1 \nu_1} 
\eta_{\mu_2 \nu_2 \mu_3 \nu_3} + \eta_{\mu_1 \nu_2} \eta_{\mu_2 \nu_3 \mu_3 \nu_1}
+\eta_{\mu_1 \nu_3} \eta_{\mu_2 \nu_1 \mu_3 \nu_2}\, .
\label{A2}
\end{align}
For the convenience in the argument later, we now solve Eq.~(\ref{e3}) with respect to 
$\ddot{h}_{ij}$.
Because the inverse of the coefficient matrix $A_{ij,kl} \equiv \eta_{(ij)kl}$ in (\ref{e3}) is given by 
\begin{align}
{A^{-1}}_{kl,mn} = -\eta_{m(k} \eta_{l) n} + \frac12 \eta_{kl} \eta_{mn} \, ,
\label{AAA1}
\end{align}
$\ddot{h}_{ij}$ can be expressed by using the terms which do not contain the second 
order derivative with respect to time, as follows, 
\begin{align}
0= \left(-\eta_{m(i} \eta_{j) n} + \frac12 \eta_{ij} \eta_{mn} \right)E_{ij} = \ddot{h}_{mn} 
+ (\text{terms without } \ddot{h})\, . 
\label{e5}
\end{align} 
In order to count the degrees of freedom of this system (that is, the number of independent initial 
values in classical mechanics), we consider the conditions that the equations in (\ref{e2}), 
which only contains the terms including the first order derivative with respect to time, 
is conserved, that is, the equations are invariant under the translation of time. 
Then we may regard $h_{0\mu}$ as dynamical fields in the equations obtained from the 
condition.
Because the obtained equations guarantee the conservation of the original equations 
(\ref{e2}), we may forget the equations except for the initial conditions. 
Thus the original equations can be regarded as the ``constraints'' on the initial values.
We continue this procedure until the conditions become the second order differential 
equations of $h_{0\mu}$ with respect to time.

Now, we regard the equations (\ref{e2}) as primary constraints:
\begin{align}
0 \approx E_{0\nu} \equiv \phi^{(1)}_\nu\, .
\label{A3}
\end{align}
Here $\approx$ means equivalence up to constraints, and the constraint (\ref{A3}) does 
not guarantee that the equation is invariant under the time evolution. 
 From the conservation of $E_{0\mu} = \phi^{(1)}_\mu $, by using the equations (\ref{e3}) 
and the constraints (\ref{e2}), we obtain
\begin{align}
0= -\dot{\phi}^{(1)}_\nu \approx m^2 \eta_{(\mu \nu ) \mu_1 \nu_1} \partial^\mu 
h^{\mu_1 \nu_1} -\mu \eta_{(\mu \nu ) \mu_1 \nu_1 \mu_2 \nu_2} \partial^\mu 
h^{\mu_1 \nu_1} \cdot h^{\mu_2 \nu_2} 
 =\partial^\mu E_{\mu \nu} \equiv \phi^{(2)}_\nu \, .
\label{A4}
\end{align}
The derivation of the above equation is a little bit tedious but if we use the equation, 
\begin{align}
\partial^\mu E_{\mu \nu} =m^2 \eta_{(\mu \nu ) \mu_1 \nu_1} \partial^\mu 
h^{\mu_1 \nu_1} -\mu \eta_{(\mu \nu ) \mu_1 \nu_1 \mu_2 \nu_2} \partial^\mu 
h^{\mu_1 \nu_1} \cdot h^{\mu_2 \nu_2}\, ,
\label{A5}
\end{align}
from the beginning (indeed the above equation (\ref{A5}) can be obtained trivially by using 
the anti-symmetry of $\eta$ tensor), we find
\begin{align}
 - E_{i\nu , i} + m^2 \eta_{(\mu \nu ) \mu_1 \nu_1} \partial^\mu h^{\mu_1 \nu_1} 
 -\mu \eta_{(\mu \nu ) \mu_1 \nu_1 \mu_2 \nu_2} \partial^\mu h^{\mu_1 \nu_1} \cdot 
h^{\mu_2 \nu_2}= -\dot{E}_{0\nu} = - \dot{\phi}^{(1)}_\nu\, .
\label{A6}
\end{align}
The terms $E_{i\nu ,i}$ can be ignored up to the equations (\ref{e3}) and the constraints 
(\ref{e2}).
Thus by using the conservation of $\phi^{(1)}_\nu$, we obtain the conditions 
$\phi^{(2)}_\nu = 0$, which is identical with (\ref{A4}) without tedious calculations. 
Now, the primary constraints (\ref{e2}) can be regarded as the condition only on the 
initial values.
Instead of primary constraints, we can impose the condition $\phi^{(2)}_\nu = 0$ for any 
time.
The primary constraints hold automatically thanks to the conditions for the conservation 
of the constraints in time $\phi^{(2)}_\nu = 0$.
Because, however, $\phi^{(2)}_\nu =0$ is the equation including only the first order 
differential equation with respect to the time, 
we have to change the equations in (\ref{A4}), which is defined by the strong equality 
 $=$ to the equations defined by the weak equality $\approx$ and we regard 
$\phi^{(2)}_\nu$ as secondary constraints $\phi^{(2)}_\nu \approx 0$.
In order to derive the conditions for the conservation of the constraints, we use one 
more relation:
\begin{align}
0&=\partial^\mu \partial^\nu E_{\mu \nu} + \frac{m^2}{2} \eta^{\mu \nu}E_{\mu \nu} 
-\mu h^{\mu \nu} E_{\mu \nu} \nn
&=-\frac{3\mu m^2}{2} \eta_{\mu \nu \mu_1 \nu_1} h^{\mu \nu} h^{\mu_1 \nu_1} 
+ \frac{\mu^2}{2} \eta_{\mu \nu \mu_1 \nu_1 \mu_2 \nu_2} h^{\mu \nu} h^{\mu_1 \nu_1} 
h^{\mu_2 \nu_2}
+\frac{3m^4}{2} h -\mu \eta_{\mu \nu \mu_1 \nu_1 \mu_2 \nu_2} \partial^\mu 
h^{\mu_1 \nu_1 } \partial^\nu h_{\mu_2 \nu_2}\, . 
\label{A7}
\end{align}
Then, by using the conservation of $\phi^{(2)}_0$, we obtain 
\begin{align}
0 \approx& - \dot{\phi}^{(2)}_0 = -\partial_0 \partial^\mu E_{\mu0} \nn
\approx& \partial^\mu \partial^\nu E_{\mu \nu} + \frac{m^2}{2} 
\eta^{\mu \nu}E_{\mu \nu} -\mu h^{\mu \nu} E_{\mu \nu} \nn
= & -\frac{3\mu m^2}{2} \eta_{\mu \nu \mu_1 \nu_1} h^{\mu \nu} h^{\mu_1 \nu_1} 
+ \frac{\mu^2}{2} \eta_{\mu \nu \mu_1 \nu_1 \mu_2 \nu_2} h^{\mu \nu} h^{\mu_1 \nu_1} 
h^{\mu_2 \nu_2} +\frac{3m^4}{2} h -\mu \eta_{\mu \nu \mu_1 \nu_1 \mu_2 \nu_2} \partial^\mu 
h^{\mu_1 \nu_1 } \partial^\nu h_{\mu_2 \nu_2}  \equiv \phi^{(3)}\, .
\label{A8}
\end{align}
Note that Eq.~(\ref{A8}) does not contain any derivative of $h_{00}$ and the second order 
derivatives of $h_{0i}$ and $h_{ij}$ with respect to time. 
On the other hand, from the conservation of the constraints $\phi^{(2)}_i$, we obtain the 
second order derivative equations for $h_{0i}$ up to the equation (\ref{e3}), 
\begin{align}
\dot{\phi}^{(2)}_i = (m^2 \eta_{ij} - \mu \eta_{ijkl}h_{kl} )\ddot{h}_{0j} 
+(\text{terms without } \ddot{h}) =0\, .
\label{A9}
\end{align}
Therefore, we can use the equations in (\ref{A9}) as the equations that describe dynamics 
of $h_{0i} $. 
Except the special configurations of fields where the matrix 
$M_{ij} = m^2 \eta_{ij} - \mu \eta_{ijkl}h_{kl}$ has any vanishing eigenvalue, we can solve 
the equations (\ref{A9}) with respect to $\ddot{h}_{0i}$ as follows, 
\begin{align}
0= \frac{1}{m^2} \left[ \eta_{ij} +\sum_{n=1}^\infty (\mathbf{H}^n)_{ij} \right] 
\dot{\phi}^{(2)}_j = \ddot{h}_{0i} + \left( \text{terms without } \ddot h \right) \, . 
\label{e6}
\end{align}
Here, $(\mathbf{H}^n)_{ij} $ is defined by 
\begin{align}
\left( \mathbf{H}^n \right)_{ij} \equiv H_{ik_1} H_{k_1 k_2} \cdots H_{k_{n-1} j}\, , \quad 
H_{ij} \equiv \frac{\mu}{m^2} \eta_{ijkl}h_{kl} \, .
\label{A10}
\end{align}
Now, let us consider the condition for the conservation of the constraint $\phi^{(3)}$.
Because $\phi^{(3)}$ does not contain any derivative of $h_{00}$ nor the second order 
derivative of $h_{0i}$, $h_{ij}$ with respect to time, $\dot{\phi}^{(3)}$ contain the first order 
derivative of $h_{00}$ and the second order derivative of $h_{0i}$ and $h_{ij}$ with respect to 
time. 
As shown, $\ddot{h}_{ij}$ and $\ddot{h}_{0i}$ can be eliminated by using Eqs.~(\ref{e5}) and 
 (\ref{e6}).
Therefore we obtain one more constraint which does not contain the terms including the  
second order derivative with respect ot time, 
\begin{align}
\dot{\phi}^{(3)} = (\text{terms without } \ddot{h}) \equiv \phi^{(4)} \approx 0\, .
\label{A11}
\end{align}
Although we do not give explicit form of this constraint, we can see that the condition 
for the conservation of the constraint (\ref{A11}) does not give any more constraints, 
which can be found as follows. 
By focusing only on the linear terms, we found the condition for the conservation of the 
constraint $\phi^{(4)}$ is given by 
\begin{align}
0=\dot{\phi}^{(4)} = \frac{3m^4}{2}\ddot{h} + \mathcal{O}(h^2)\, .
\label{A12}
\end{align}
We should note that the first term cannot be canceled by $\mathcal{O}(h^2)$ terms.
In addition, because the first term contains $\ddot{h}_{00}$, this equation defines  the 
dynamics of $h_{00}$.
Therefore the condition for the conservation of $\phi^{(4)}$ does not give any more 
constraints.
Finally, we obtain 10 equations including second order derivative with respect to time 
which describe the dynamics of $h_{\mu \nu}$ and 10 constraints which restrict the initial 
values.
Then we find the pseudo-linear theory has $(20-10)/2 = 5$ degrees of freedom on the 
flat space.

\section{Pseudo-linear theory on curved space}

Before the discussion of the new model on curved space-time, let us briefly 
review on the Fierz-Pauli theory on curved space-time. In \cite{Buchbinder:1999ar}, 
Buchbinder et al. showed that the Fierz-Pauli theory on the non-trivial background must have 
the non-minimal coupling terms and the space-time 
is required to be the Einstein manifold in order to keep the consistency. This ghost-free action is given by
\newpage
\begin{align}
S=& \int d^4x \sqrt{-g}\left\{ \frac12 \nabla_\mu h \nabla^\mu h 
 -\frac12 \nabla_\mu h_{\nu \rho} \nabla^\mu h^{\nu \rho} 
 - \nabla^\mu h_{\mu \nu} \nabla^\nu h + \nabla_\mu h_{\nu \rho } 
\nabla^\rho h^{\nu \mu}+\frac{m^2}{2} g^{\mu_1 \nu_1 \mu_2 \nu_2 } h_{\mu_1 \nu_1} 
h_{\mu_2 \nu_2} \right. \nn
& \left. 
+\frac{\xi}{4} R h_{\alpha \beta} h^{\alpha \beta} +\frac{1-2\xi}{8} R h^2
\right\}\, , 
\label{CFP}
\end{align}
This suggests that these non-minimal coupling terms should be added to the new model 
(\ref{hh10}) on curved space-time. Thus, 
we consider the following action.
\begin{align}
S=& \int d^4x \sqrt{-g}\left\{ \frac12 \nabla_\mu h \nabla^\mu h 
 -\frac12 \nabla_\mu h_{\nu \rho} \nabla^\mu h^{\nu \rho} 
 - \nabla^\mu h_{\mu \nu} \nabla^\nu h + \nabla_\mu h_{\nu \rho } 
\nabla^\rho h^{\nu \mu}+\frac{m^2}{2} g^{\mu_1 \nu_1 \mu_2 \nu_2 } h_{\mu_1 \nu_1} 
h_{\mu_2 \nu_2} \right. \nn
& \left. -\frac{\mu}{ 3!} g^{\mu_1 \nu_1 \mu_2 \nu_2 \mu_3 \nu_3 } h_{\mu_1 \nu_1} 
h_{\mu_2 \nu_2} h_{\mu_3 \nu_3} -\frac{\lambda}{ 4!} 
g^{\mu_1 \nu_1 \mu_2 \nu_2 \mu_3 \nu_3 \mu_4 \nu_4} h_{\mu_1 \nu_1} 
h_{\mu_2 \nu_2} h_{\mu_3 \nu_3} h_{\mu_4 \nu_4}
+\frac{\xi}{4} R h_{\alpha \beta} h^{\alpha \beta} +\frac{1-2\xi}{8} R h^2
\right\}\, , 
\label{c7}
\end{align}
Here the metric $g$ is chosen to be the Einstein manifold, where the curvatures satisfy the 
following condition: 
\begin{align}
R_{\mu \nu} = \frac{R}{4} g_{\mu \nu}\, . 
\label{c8}
\end{align}
As a first step, let us consider the  $\lambda=0$ case and prove that there appear 5 degrees of freedom. 
By using the action (\ref{c7}) with the $\lambda=0$, we find that $h_{\mu \nu}$ obeys the following equations 
\begin{align}
0=E_{\mu \nu} = &g^{\alpha \beta} \nabla_\alpha \nabla_\beta h_{\mu \nu} 
 -g_{\mu \nu} g^{\alpha \beta} g^{\gamma \delta}
\nabla_\alpha \nabla_\beta h_{\gamma \delta} 
+ g_{\mu \nu} g^{\alpha \gamma } g^{\beta \delta} \nabla_\alpha \nabla_\beta
h_{\gamma \delta} -2 g^{\sigma \rho} \nabla_\sigma  \nabla_{( \mu} h_{\nu ) \rho} 
+ g^{\alpha \beta} \nabla_\mu \nabla_\nu h_{\alpha \beta} \nn
&+ m^2 {g_{(\mu \nu )}}^{\alpha \beta} h_{\alpha \beta}
+\frac{\xi}{2} R h_{\mu \nu} + \frac{1-2\xi}{4} R g^{\alpha \beta} g_{\mu \nu} 
h_{\alpha \beta} -\frac{\mu}{2} {g_{(\mu \nu )}}^{\mu_1 \nu_1 \mu_2 \nu_2} 
h_{\mu_1 \nu_1} h_{\mu_2 \nu_2} \nn
=&-{g_{(\mu \nu ) }}^{\mu_1 \nu_1 \mu_2 \nu_2} \nabla_{\mu_1} \nabla_{\nu_1} 
h_{\mu_2 \nu_2} +(\text{terms without } \nabla \nabla h) \nn
=& -g_{i (\mu } g_{\nu) j} g^{ij 00 \mu_2 \nu_2} \nabla_0 \nabla_0 h_{\mu_2 \nu_2} 
+ (\text{terms without } \nabla_0 \nabla_0 h)\, . 
\label{c1}
\end{align}
In this section, we regard the equations which do not include $\nabla_0 \nabla_0 h$
 (or $\partial_0 \partial_0 h$) as constraints.
Because $E_{\mu \nu}$ can be expressed as in the last line of (\ref{c1}), we find 
that $E_{0\mu}$ contain the second order derivative terms with respect to time. 
We should note, however, that these terms including the second order derivatives with 
respect to time can be eliminated by the linear combinations of $E_{\mu \nu}$ as follows, 
\begin{align}
{E^0}_{\nu} &= g^{00} E_{0\nu} + g^{0i} E_{i\nu} \nn
&= -g_{\nu \sigma} g^{(0 \sigma ) \mu_1 \nu_1 \mu_2 \nu_2} \nabla_{\mu_1} \nabla_{\nu_1} 
h_{\mu_2 \nu_2}+(\text{terms without } \nabla \nabla h) \nn
&= (\text{terms without } \nabla_0 \nabla_0 h) \equiv \phi^{(1)}_\nu \approx 0\, .
\label{A13}
\end{align}
Therefore, $\phi^{(1)}_\nu \equiv {E^0}_\nu$ can be regarded as primary constraints.
Then $(ij)$-components of Eq.~(\ref{c1}) have the following forms:
\begin{align}
0=E_{ij} = -g_{m(i} g_{j)n} g^{mn 00 kl} \nabla_0 \nabla_0 h_{kl} 
+(\text{terms without } \nabla_0 \nabla_0 h)\, .
\label{c2}
\end{align}
In order to solve Eq.~(\ref{c2}) with respect to $\nabla_0 \nabla_0 h_{ij}$, we use the ADM 
decomposition: 
\begin{align}
g^{00} = -\frac{1}{N^2} \, , \quad & g_{0k}=N_k\, , \quad g_{ij}=e_{ij}\, , \nn
g_{00}=N^k N_k-N^2\, , \quad & g^{0i}=\frac{N^i}{N^2}\, , \quad 
g^{ij}=e^{ij} -\frac{N^iN^j}{N^2}\, .
\label{A14}
\end{align}
Here $e_{ij}$ is a three dimensional metric field and has the following properties, 
\begin{align}
e^{ij}e_{ij} = {\delta^i}_j\, , \quad N^i\equiv e^{ij}N_j\, , \quad 
e^{ij}=g^{ij}-\frac{g^{0i}g^{0j}}{g^{00}}\, .
\label{A15}
\end{align}
By using the ADM decomposition, the coefficient matrix in equations (\ref{c2}) can be 
expressed as (see (\ref{p2}) in Appendix), 
\begin{align}
{A_{ij}}^{,kl} \equiv -g_{m(i} g_{j)n} g^{mn 00 kl} = \frac{1}{N^2} {e_{(ij)}}^{kl} \, .
\label{c3}
\end{align}
Here 
\begin{align}
e_{i_1 j_1 i_2 j_2 \cdots i_n j_n} \equiv e_{i_1 j_1}e_{ i_2 j_2} \cdots e_{ i_n j_n} 
 -e_{i_1 j_2}e_{ i_2 j_1} \cdots e_{ i_n j_n} + \cdots \, .
\label{A16}
\end{align}
We now define the raising and lowering the indices in $e_{i_1j_1 \cdots i_n j_n}$ by 
using $e^{ij}$ and $e_{ij}$.
The inverse matrix of the coefficient matrix (\ref{c3}) is expressed as
\begin{align}
{{A^{-1}}_{kl}}^{,mn} =N^2 \left( \frac12 e_{kl}e^{mn} -{e_{(k}}^m {e_{l)}}^n  \right) 
&=\frac{1}{g^{00}} \left\{  {g_{(k}}^m {g_{l)}}^n -\frac12 g_{kl}(g^{mn} 
 -\frac{g^{0m}g^{0n}}{g^{00}}) \right\}\, , \nn
{A_{ij}}^{,kl}{{A^{-1}}_{kl}}^{,mn}&=\delta_{(i}^k \delta_{j)}^l\, . 
\label{c20}
\end{align}
Then, we can solve Eq.~(\ref{c2}) with respect to $\nabla_0 \nabla_0 h_{ij}$ as follows, 
\begin{align}
0= \frac{1}{g^{00}} \left\{  {g_{(k}}^i {g_{l)}}^j -\frac12 g_{kl} \left( g^{ij} 
 -\frac{g^{0i}g^{0j}}{g^{00}} \right) \right\} E_{ij}
=\nabla_0 \nabla_0 h_{ij} + (\text{terms without } \nabla_0 \nabla_0 h) \, . 
\label{c4}
\end{align}
Because Eq.~(\ref{c2}) gives 6 independent equations including the second order 
derivative with respect to time and these equations are also independent of the primary 
constraints $\phi^{(1)}_\nu$, these equations describe the dynamics of $h_{ij}$. 
In order to obtain the conditions for the conservations of the primary constraints, we use 
the following relations:
\begin{align}
\nabla^\mu E_{\mu \nu} =& \frac{R}{4}g^{\alpha \beta} \nabla_\nu  h_{\alpha \beta}
 -\frac{R}{2} g^{\sigma \rho} \nabla_{\sigma} h_{\rho \nu} 
+ m^2 g_{\nu \nu_1} g^{\mu_1 \nu_1 \mu_2 \nu_2} \nabla_{\mu_1} h_{\mu_2 \nu_2} 
\nn
& +\frac{\xi}{2}  R g^{\sigma \rho} \nabla_\sigma h_{\rho \nu}
+ \frac{1-2\xi}{4} R g^{\alpha \beta} \nabla_\nu h_{\alpha \beta}
-\mu g_{\nu \nu_1} g^{( \mu_1 \nu_1 ) \mu_2 \nu_2 \mu_3 \nu_3} \left( \nabla_{\mu_1} h_{\mu_1 \nu_1} \right) h_{\mu_2 \nu_2} 
\nn
&= \left( \frac{1-\xi}{2} R + m^2 \right) g_{\nu \nu_1} 
g^{\mu_1 \nu_1 \mu_2 \nu_2} \nabla_{\mu_1} h_{\mu_2 \nu_2} 
 -\mu g_{\nu \nu_1} g^{(\mu_1 \nu_1 ) \mu_2 \nu_2 \mu_3 \nu_3} 
\left( \nabla_{\mu_1} h_{\mu_2 \nu_2} \right) h_{\mu_3 \nu_3}\, .
\label{A17}
\end{align}
Then the secondary constraints are obtained as 
\begin{align}
\partial_0 \phi^{(1)}_\nu = \partial_0 {E^0}_\nu \approx \nabla^\mu E_{\mu \nu} 
\equiv \phi^{(2)}_\nu \approx 0\, .  
\label{A18}
\end{align}
For convenience, we choose independent constraints as follows,  
\begin{align}
\phi^{(2)0} \equiv g^{00} \phi^{(2)}_0 + g^{0i} \phi^{(2)}_i \approx 0\, , \quad 
\phi^{(2)}_i \approx 0\, .
\label{A19}
\end{align}
Furthermore, by using the following relation:
\begin{align}
\nabla^\mu &\nabla^\nu E_{\mu \nu} +\frac{m^2}{2} g^{\mu \nu}E_{\mu \nu}  
 -\mu h^{\mu \nu} E_{\mu \nu} +\frac{1-\xi }{4} R g^{\mu \nu} E_{\mu \nu} \nn
=&h \left( \frac{3m^4}{2} + \frac{5-6\xi}{4}m^2 R +\frac{(1-\xi )(2-3\xi )}{8} R^2 
\right) \nn
&-\frac{3\mu m^2}{2} g^{\mu_1 \nu_1 \mu_2 \nu_2} h_{\mu_1 \nu_1} h_{\mu_2 \nu_2}
 -\mu g^{\mu_1 \nu_1 \mu_2 \nu_2 \mu_3 \nu_3 } 
\left( \nabla_{\mu_1} h_{\mu_2 \nu_2} \right) \nabla_{\nu_1} h_{\mu_3 \nu_3} \nn
&+\frac{\mu^2}{2} g^{\mu_1 \nu_1 \mu_2 \nu_2 \mu_3 \nu_3 } h_{\mu_1 \nu_1 } 
h_{\mu_2 \nu_2} h_{\mu_3 \nu_3} - \frac{7-9\xi}{12} \mu R g^{\mu_1 \nu_1 \mu_2 \nu_2} 
h_{\mu_1 \nu_1} h_{\mu_2 \nu_2} -\mu C^{\mu \alpha \nu \beta} h_{\mu \nu} 
h_{\alpha \beta}\, ,
\label{A19B}
\end{align}
we find one more constraint:
\begin{align}
\partial_0 \phi^{(2)0} \approx \nabla^\mu \nabla^\nu E_{\mu \nu} 
+\frac{m^2}{2} g^{\mu \nu}E_{\mu \nu}   -\mu h^{\mu \nu} E_{\mu \nu} 
+\frac{1-\xi }{4} R g^{\mu \nu} E_{\mu \nu}
\equiv \phi^{(3)} \approx 0\, .
\label{A20}
\end{align}

We have to note that the non-minimal coupling terms in (\ref{A17}) play a very important role for the existence of the constraint $\phi^{(3)}$.
The term $\frac{R}{4}g^{\alpha \beta} \nabla_\nu  h_{\alpha \beta}-\frac{R}{2} g^{\sigma \rho} \nabla_{\sigma} h_{\rho \nu}$ in (\ref{A17})
contains the derivatives of $h_{00}$ with respect to time and this prevents us from having the appropriate number of constraints. 
The contribution from the non-minimal couplings, 
however, cancels out these time-derivative terms and enables the system to have 5 degrees of freedom.

On the other hand, the term including the curvature tensor does not appear from the cubic potential in $\nabla^{\mu} E_{\mu \nu}$ and, 
as a result, does not contain any derivative of $h_{00}$ with respect to time. 
Needless to say, the derivative of $h_{00}$ with respect to time emerges when
we act another covariant derivative on $\nabla^{\mu}E_{\mu \nu}$, but this term is also eliminated by the term $h^{\mu \nu}E_{\mu \nu}$ in (\ref{A20}).
This means that we do not need any additional non-minimal coupling terms for this system to be ghost-free. Generally, we can add a new non-minimal coupling term
$Rg^{\mu_1 \nu_1 \mu_2 \nu_2 \mu_3 \nu_3} h_{\mu_1 \nu_1 } h_{\mu_2 \nu_2} h_{\mu_3 \nu_3}$ without any ghost, but this fact does not change the following 
analysis. This is because the scalar curvature is constant on the Einstein manifold and the effect of the new term can be absorbed by the redefinition of
$\mu$.
On the other hand, by using the ADM decomposition, the conditions for the conservation 
of $\phi^{(2)}_i$ have the following forms:
\begin{align}
& \partial_0 \phi^{(2)}_i \approx \nabla_0 \nabla^\mu E_{\mu i} 
= {B_i}^j \nabla_0 \nabla_0 h_{0j} + {C_{i}}^{kl} \nabla_0 \nabla_0 h_{kl} 
+ (\text{terms without } \nabla_0 \nabla_0 h)=0\, , \nn
& {B_i}^j \equiv \frac{1}{N^2} \left[ \left( \frac{1-\xi}{2} R +m^2 \right) {\delta_i}^j 
 -\mu e^{j~mn}_{~i}h_{mn}  \right]\, , \nn
& {C_i}^{kl} \equiv -\frac{1}{N^2} \left[ \left( \frac{1-\xi}{2} R +m^2 \right) N^k {\delta_i}^l 
 +\mu \left\{ {{e^{kl}}_{i}}^j h_{0j} +N^k {e_{i}}^{lmn} h_{mn} +N^m {e_i}^{nkl}h_{mn} \right\}  
\right]\, .
\label{A21}
\end{align}
Here we have used (\ref{p1}) and (\ref{p2}). 
Then $\nabla_0 \nabla_0 h_{kl}$ can be eliminated by using Eq.~(\ref{c4}) as follows, 
\begin{align}
\partial_0 \phi^{(2)}_i \approx \nabla_0 \nabla^\mu E_{\mu i} 
 - {C_i}^{kl} {{A^{-1}}_{kl}}^{,mn} E_{mn} 
= {B_i}^a \nabla_0 \nabla_0 h_{0a} + (\text{terms without} \nabla_0 \nabla_0 h)=0\, .
\label{A22}
\end{align}
Thus, we find that the conditions (\ref{A21}) describe the dynamics of $h_{0i}$ 
so that the constraints in (\ref{A19B}) are conserved. 
Except the special configuration of field where the matrix ${B_i}^j$ has vanishing 
eigenvalue,  we can solve the equations in (\ref{A21}) with respect to 
$\nabla_0 \nabla_0 h_{0i}$ as follows,
\begin{align}
&{{B^{-1}}_k}^i \left[ \nabla_0 \nabla^\mu E_{\mu i} 
 - {C_i}^{kl} {{A^{-1}}_{kl}}^{,mn} E_{mn} \right] 
= \nabla_0 \nabla_0 h_{0k}  + (\text{terms without }\nabla_0 \nabla_0 h) =0\, , \nn
& {{B^{-1}}_i}^j = \frac{N^2}{\frac{1-\xi}{2} R +m^2} \left[ {\delta_i}^j 
+ \sum_{n=1}^\infty {(\mathbf{H}^n)_i}^j \right]\, , \nn
& {(\mathbf{H}^n)_i}^j \equiv H_{ik_1} e^{k_1 l_1} H_{l_1 k_2} e^{k_2 l_2} 
\cdots H_{l_{n-1} k_n} e^{k_n j}\, , \quad 
H_{ij} \equiv \frac{\mu}{\frac{1-\xi}{2} R +m^2} e^{~~mn}_{ij}h_{mn}\, .
\label{A23}
\end{align}
As in the case of the flat background, the constraint obtained from the condition for 
the conservation of $\phi^{(3)}$ has the following form:
\begin{align}
\partial_0 \phi^{(3)} \approx (\text{terms without }\nabla_0 \nabla_0 h ) \equiv 
\phi^{(4)} \approx 0\, .
\label{A24}
\end{align}
By focusing the linear terms, we find that the condition for the conservation of the 
constraint $\phi^{(4)}$ defines the dynamics of $h_{00}$. 
As a result, the pseudo-linear theory described by action (\ref{c7}) has 5 degrees of 
freedom on the Einstein manifold (\ref{c8}).

\section{$\lambda \neq 0 $ case}

We now investigate if the arguments in the previous sections can be extended to the  
$\lambda \neq 0$ case.
In the $\lambda \neq 0$ case,  the equations are modified as follows,
\begin{align}
0=E_{\mu \nu} = &g^{\alpha \beta} \nabla_\alpha \nabla_\beta h_{\mu \nu} 
 -g_{\mu \nu} g^{\alpha \beta} g^{\gamma \delta}
\nabla_\alpha \nabla_\beta h_{\gamma \delta} 
+ g_{\mu \nu} g^{\alpha \gamma } g^{\beta \delta} \nabla_\alpha \nabla_\beta
h_{\gamma \delta} -2 g^{\sigma \rho} \nabla_\sigma  \nabla_{( \mu} h_{\nu ) \rho} \nn 
&+ g^{\alpha \beta} \nabla_\mu \nabla_\nu h_{\alpha \beta}
+ m^2 {g_{(\mu \nu )}}^{\alpha \beta} h_{\alpha \beta}
+\frac{\xi}{2} R h_{\mu \nu} + \frac{1-2\xi}{4} R g^{\alpha \beta} g_{\mu \nu} 
h_{\alpha \beta} \nn
&-\frac{\mu}{2} {g_{(\mu \nu )}}^{\mu_1 \nu_1 \mu_2 \nu_2} h_{\mu_1 \nu_1} 
h_{\mu_2 \nu_2} -\frac{\lambda}{3!} 
{g_{(\mu \nu )}}^{ \mu_1 \nu_1 \mu_2 \nu_2 \mu_3 \nu_3} h_{\mu_1 \nu_1} h_{\mu_2 \nu_2} 
h_{\mu_3 \nu_3}\, .
\label{A25}
\end{align}
Then we obtain the following primary constraint:
\begin{align}
{E^0}_{\nu} \equiv \phi^{(1)}_\nu \approx 0\, .
\label{A26}
\end{align}
By using the conservation of the constraint $\phi^{(1)}_\nu$ (\ref{A26}), we find that the 
secondary constraints are given by
\begin{align}
\nabla^\mu E_{\mu \nu} =& \left( \frac{1-\xi}{2} R + m^2 \right) g_{\nu \nu_1} 
g^{\mu_1 \nu_1 \mu_2 \nu_2} \nabla_{\mu_1} h_{\mu_2 \nu_2} 
 -\mu g_{\nu \nu_1} g^{(\mu_1 \nu_1 ) \mu_2 \nu_2 \mu_3 \nu_3} 
\left( \nabla_{\mu_1} h_{\mu_2 \nu_2} \right) h_{\mu_3 \nu_3} \nn
&-\frac{\lambda}{2} g_{\nu \nu_1} 
g^{( \mu_1 \nu_1 ) \mu_2 \nu_2 \mu_3 \nu_3 \mu_4 \nu_4} 
\left( \nabla_{\mu_1} h_{\mu_2 \nu_2} \right) h_{\mu_3 \nu_3} h_{\mu_4 \nu_4} 
\equiv \phi^{(2)}_\nu \approx 0\, .
\label{A27}
\end{align}
As in the $\lambda =0$ case, the conservation of the constraints $\phi^{(2)}_i$ gives 
3 equations describing the dynamics of $h_{0i}$.
The equations of the conservation have the following structure, 
\begin{align}
& {B_i}^j \nabla_0 \nabla_0 h_{j0}
+ (\text{terms without } \nabla_0 \nabla_0 h) =0 \, , \nn
&{B_i}^j \equiv \frac{1}{N^2}\left[ \left( \frac{1-\xi}{2} R +m^2  \right) {\delta_i}^j 
 -\mu e^{j~mn}_{~i}h_{mn} 
 -\frac{\lambda}{2} e_{ii_1}e^{(i_1 j) i_2 j_2 i_3 j_3} h_{i_2 j_2} h_{i_3 j_3}  \right] \, .
\label{A28}
\end{align}
The matirix ${B_i}^j$, which is the coefficient of $\nabla_0 \nabla_0 h_{j0}$ can be 
eliminated by using the following inverse matrix:
\begin{align}
&{{B^{-1}}_i}^j =\frac{N^2}{\frac{1-\xi}{2} R +m^2} \left[ {\delta_i}^j + \sum_{n=1}^\infty 
{(\mathbf{H}^n)_i}^j \right] \, , \nn
& {(\mathbf{H}^n)_i}^j \equiv H_{ik_1} e^{k_1 l_1} H_{l_1 k_2} e^{k_2 l_2} \cdots H_{l_{n-1} k_n} 
e^{k_n j}\, , \quad 
H_{ij} \equiv \frac{1}{\frac{1-\xi}{2} R +m^2} \left[ \mu e^{~~mn}_{ij} h_{mn} 
+ \frac{\lambda}{2} e_{(ij)}^{~~~~kl mn}h_{kl} h_{mn} \right] \, .
\label{A29}
\end{align}
Furthermore the conservation of $\phi^{(2)0}$ gives an additional constraint:
\begin{align}
\nabla^\mu &\nabla^\nu E_{\mu \nu} +\frac{m^2}{2} g^{\mu \nu}E_{\mu \nu}   
+\frac{1-\xi }{4} R g^{\mu \nu} E_{\mu \nu}-\mu h^{\mu \nu} E_{\mu \nu}
 +\frac{\lambda}{2} g^{00 ij kl mn} {A^{-1}_{ij}}^{,ab} E_{ab} h_{kl} h_{mn} \nn
 =&-\mu g^{(\mu_1 \nu_1 ) \mu_2 \nu_2 \mu_3 \nu_3} 
\left( \nabla_{\mu_1} h_{\mu_2 \nu_2} \right) \nabla_{\nu_1} h_{\mu_3 \nu_3} - \lambda 
g^{(\mu_1 \nu_1 ) \mu_2 \nu_2 \mu_3 \nu_3 \mu_4 \nu_4} 
\left( \nabla_{\mu_1} h_{\mu_2 \nu_2} \right) \left( \nabla_{\nu_1} h_{\mu_3 \nu_3} \right) 
h_{\mu_4 \nu_4} \nn
& -\lambda g^{(0i)  \mu_2 \nu_2 \mu_3 \nu_3 \mu_4 \nu_4} 
\left( \nabla_0 \nabla_i h_{\mu_2 \nu_2}\right) h_{\mu_3 \nu_3}  h_{\mu_4 \nu_4}
+ \frac{\lambda}{2} g^{00 ij kl mn} {A^{-1}_{ij}}^{,ab} 
\left(-2 {g_{ab}}^{(0c) \mu \nu} \nabla_0 \nabla_c h_{\mu \nu} \right) h_{kl} h_{mn} \nn
&+ (\text{terms without any time derivatives of } h )\equiv \phi^{(3)} \approx 0\, . 
\label{l1}    
\end{align}
Here ${A^{-1}_{ij}}^{,kl}$ is defined by (\ref{c20}). 
By using the expression (\ref{l1}), we find that the derivative of $\phi^{(3)}$ with respect 
to time does not contain the second order derivatives of $h_{00}$ with respect to time 
but the second order derivatives of $h_{0i}$ and  $h_{ij}$ with respect to time.
Since the second order derivatives of $h_{0i}$ and $h_{ij}$ with respect to time can be 
eliminated as in the $\lambda =0$ case, there appears one more constraint:
\begin{align}
\phi^{(4)} \approx 0\, .
\label{A30}
\end{align}
Therefore, even in the case of $\lambda \neq 0$, the pseudo-linear theory has 5 degrees 
of freedom on the Einstein manifold.

\section{A new non-minimal coupling term}

In \cite{Buchbinder:1999ar}, in order to eliminate a ghost, non-minimal coupling terms were 
added to the Fierz-Pauli action. 
In this section, we show that there is another kind of non-minimal coupling which does not 
induce the ghost. 
We should note that the constraint $\phi^{(3)}$ is essential to exclude the extra degrees of 
freedom. 

We now assume the quadratic part in the action to be more general form than that in 
\cite{Buchbinder:1999ar} on the Einstein manifold, as follows, 
\begin{align}
S= \int d^4 x \sqrt{-g} & \left[ \frac12 g^{\mu_1 \nu_1 \mu_2 \nu_2 \mu_3 \nu_3 } 
\nabla_{\mu_1} h_{\mu_2 \nu_2} \nabla_{\nu_1} h_{\mu_3 \nu_3} 
+ \frac{m^2}{2} g^{\mu_1 \nu_1 \mu_2 \nu_2} h_{\mu_1 \nu_1} h_{\mu_2 \nu_2} \right. \nn
& \left. + \frac{\alpha}{2} R h^2 + \frac{\beta}{2} R h_{\mu \nu} h^{\mu \nu} 
+ \frac{\gamma}{2} C^{\mu \alpha \nu \beta}h_{\mu \nu} h_{\alpha \beta} \right] \, ,
\label{AA1}
\end{align}
and find the combinations of the parameters which do not induce the ghost. 
We should note that the kinetic term in (\ref{AA1}) is not identical with that in (\ref{CFP}) because the 
non-commutativity of the covariant derivatives induces the curvature tensor.
That is, the first term in (\ref{AA1}) is expanded as follows:
\begin{align}
\frac12 g^{\mu_1 \nu_1 \mu_2 \nu_2 \mu_3 \nu_3 } \nabla_{\mu_1} h_{\mu_2 \nu_2} \nabla_{\nu_1} h_{\mu_3 \nu_3} 
=& \frac12 \nabla_\mu h \nabla^\mu h 
 -\frac12 \nabla_\mu h_{\nu \rho} \nabla^\mu h^{\nu \rho} 
 - \nabla^\mu h_{\mu \nu} \nabla^\nu h + \nabla_\mu h_{\nu \rho} \nabla^\rho h^{\nu \mu} \nn 
& +\frac{R}{4} h_{\alpha \beta} h^{\alpha \beta} -\frac{R}{8} h^2 
 -\frac12 C^{\mu \alpha \nu \beta} h_{\mu \nu} h_{\alpha \beta} + \frac{R}{12} g^{\mu_1 \nu_1 \mu_2 \nu_2}h_{\mu_1 \nu_1} h_{\mu_2 \nu_2}
\label{rp1}
\end{align}
Here we have used the following relation between the Riemann curvature 
$R_{\mu \alpha \nu \beta}$ and the Weyl tensor $C_{\mu \alpha \nu \beta}$ 
on the Einstein manifold (\ref{c8}),
\begin{align}
R_{\mu \alpha \nu \beta} 
= C_{\mu \alpha \nu \beta} + \frac{R}{12}g_{\mu \nu \alpha \beta}\, . 
\label{d9}
\end{align}
Thus, we have to subtract the contribution from the terms including the curature tensor when we use 
$g^{\mu_1 \nu_1 \mu_2 \nu_2 \cdots}$ to express the kinetic term. In (\ref{AA1}), these extra terms are 
included in the remaining non-minimal coupling terms because, on the Einstein manifold, the Riemann 
tensor can be decomposed into the Weyl tensor and the Ricci scalar and the Ricci tensor can be expressed 
in terms of the Ricci scalar.

The contribution to the equation from the kinetic terms in the action (\ref{AA1}) 
is given by 
\begin{align}
E_K^{\mu \nu} &\equiv -g^{(\mu \nu )\mu_1 \nu_1 \mu_2 \nu_2} 
\nabla_{\mu_1} \nabla_{\nu_1} h_{\mu_2 \nu_2} \nn
&= -g^{\mu \nu \mu_1 \nu_1 \mu_2 \nu_2} \nabla_{\mu_1} \nabla_{\nu_1} h_{\mu_2 \nu_2}
+ \frac12 g^{[\mu \nu ] \mu_1 \nu_1 \mu_2 \nu_2} R_{\nu_1~\mu_1 \mu_2}^{~~\sigma} 
h_{\sigma \nu_2} \, .
\label{d1}
\end{align}
Here we have used 
\begin{align}
R_{\lambda \alpha \beta \gamma} + R_{\lambda \beta \gamma \alpha} 
+ R_{\lambda \gamma \alpha \beta} =0 \, . 
\label{d9b}
\end{align}
By using (\ref{d9})
and the following identities, 
\begin{align}
g^{\mu \nu \mu_1 \nu_1 \mu_2 \nu_2} C_{\nu_1~\mu_1 \mu_2}^{~~\sigma}
&= (g^{\mu_1 \nu_1}g^{\mu \nu \mu_2 \nu_2} +g^{\mu \nu_1} g^{\mu_2 \nu \mu_1 \nu_2}
 + g^{\mu_2 \nu_1} g^{\mu_1 \nu \mu \nu_2} ) C_{\nu_1~\mu_1 \mu_2}^{~~\sigma}  \nn
&= 2C^{\mu \sigma \nu_2 \nu} \, , \nn
g^{\mu \nu \mu_1 \nu_1 \mu_2 \nu_2}g_{\nu_1 \mu_1~\mu_2}^{~~~~~\sigma}
 &= 2g^{\mu \nu \mu_1 \nu_1 \mu_2 \nu_2} g_{\nu_1 \mu_1} g^\sigma_{~\mu_2} 
=4g^{\mu \nu \sigma \nu_2} \, ,
\label{d2}
\end{align}
we find that there is a symmetry with respect of the exchange of the indices $\mu$ and 
$\nu$ and therefore the last term in (\ref{d1}) vanishes. 
Then we obtain the following expression, 
\begin{align}
\nabla_\mu E_K^{\mu \nu} &= \frac12 g^{\mu \nu \mu_1 \nu_1 \mu_2 \nu_2} 
R_{\nu_1~\mu_1 \mu}^{~~\sigma}
\left[ \nabla_\sigma h_{\mu_2 \nu_2 } - \nabla_{\nu_2} h_{\mu_2 \sigma} \right] \nn
&=-C^{\mu \alpha \nu \beta} \nabla_\mu h_{\alpha \beta} + \frac{R}{6} 
g^{\mu \nu \alpha \beta} \nabla_\mu h_{\alpha \beta} \, .
\label{d6}
\end{align}
Here in the first line, we have used 
$g^{\mu \nu \mu_1 \nu_1 \mu_2 \nu_2} R_{\mu_2~\mu \mu_1}^{~~\sigma} =0$, 
which is obtained from (\ref{d9b}), and in the second line, we have used (\ref{d2}). 
 From the analyses in the previous sections, we know that there exists the 
constraint $\phi^{(3)}$ if the expression of $\nabla_\mu E^{\mu \nu}$ does not 
include the derivative of $h_{00}$ with respect to time although this is not the 
necessary condithion that there only exist five degrees of freedom. 
This condition is satisfied in $\nabla_\mu E_K^{\mu \nu}$ and also trivially in the 
contribution from the mass terms and therefore the model does not include ghost even if 
we set all parameters 0. 
On the other hand, we may add extra terms with non-minimal coupling if the terms 
do not induce ghost. 
This extra terms can be added if we choose $\beta= -\alpha$, 
\begin{align}
S= \int d^4 x \sqrt{-g} & \left[ \frac12 g^{\mu_1 \nu_1 \mu_2 \nu_2 \mu_3 \nu_3 } 
\nabla_{\mu_1} h_{\mu_2 \nu_2} \nabla_{\nu_1} h_{\mu_3 \nu_3} 
+ \frac{m^2}{2} g^{\mu_1 \nu_1 \mu_2 \nu_2} h_{\mu_1 \nu_1} h_{\mu_2 \nu_2} \right. \nn
& \left. + \frac{\alpha}{2} R g^{\mu_1 \nu_1 \mu_2 \nu_2} h_{\mu_1 \nu_1} h_{\mu_2 \nu_2} 
+ \frac{\gamma}{2} C^{\mu \alpha \nu \beta}h_{\mu \nu} h_{\alpha \beta} \right] \, .
 \label{d3}
\end{align}
On the Einstein manifold, because the scalar curvature is constant the terms which are 
proportional to $\alpha$ can be absorbed into the redefinition of the mass terms, which 
tells that the ghost is not generated on the Einstein manifold. 
Furthermore, the term proportional to $\gamma$ change only the coefficient of the first 
term of the second line in (\ref{d6}) and therefore this term does not induce the ghost. 

We now show that the model in (\ref{d3}) does not surely include the ghost. 
The equation given by the variation of the action (\ref{d3}) is given by 
\begin{align}
E^{\mu \nu} \equiv  -g^{(\mu \nu )\mu_1 \nu_1 \mu_2 \nu_2} \nabla_{\mu_1} \nabla_{\nu_1} 
h_{\mu_2 \nu_2} + \left(m^2+\alpha R \right) g^{\mu \nu \mu_1 \nu_1} h_{\mu_1 \nu_1} 
+ \gamma C^{\mu \alpha \nu \beta} h_{\alpha \beta} =0 \, .
\label{AA2}
\end{align}
Then we obtain the primary constraint $E^{0\mu}\equiv \phi^{(1)\mu} \approx 0$. 
The secondary constraint is given by 
\begin{align}
\nabla_\mu E^{\mu \nu} 
= (\gamma -1)C^{\mu \alpha \nu \beta} \nabla_\mu h_{\alpha \beta}
+\left\{ m^2 + R \left( \frac{1}{6} + \alpha \right) \right\}g^{\mu \nu \alpha \beta} 
\nabla_\mu h_{\alpha \beta}  
+\gamma \nabla_\mu C^{\mu \alpha \nu \beta} \cdot h_{\alpha \beta} \equiv \phi^{(2)\nu} 
\approx 0 \, . 
\label{d4}
\end{align}
The last term 
$\nabla_\mu C^{\mu \alpha \nu \beta} \cdot h_{\alpha \beta}$ 
vanishes on the Einstein manifold by the Bianchi identity. 
The second order derivative of $\phi^{(2)i}$ with respect to time does not contain the 
second order derivative of $h_{00}$ with respect to time and the constraint (\ref{d4}) 
determines the values of $\nabla_0 \nabla_0 h_{0i}$. 
We also obtain 
\begin{align}
\nabla_\mu \nabla_\nu & E^{\mu \nu} -(\gamma -1 ) C^{0 i 0 j} A^{-1~~,kl}_{~~ij}E_{kl} 
+\frac{1}{2}\left\{ m^2 +R(\frac{1}{6} + \alpha ) \right\} g^{\alpha \beta} E_{\alpha \beta} 
\nn
& =-(\gamma -1) C^{i(0j)\beta} \nabla_0 \nabla_j h_{i\beta} - (\gamma -1) 
C^{0i0j}A^{-1~~,kl}_{~~ij}(-2 g_{kl}^{~~(0c) \mu \nu}) \nabla_0 \nabla_c h_{\mu \nu} \nn
&~~
+ \left( \text{terms which do not include $\partial_0 h$} \right) 
\equiv \phi^{(3)} \approx 0 \, .
\label{AA4}
\end{align}
Here $A^{-1}$ is defined by (\ref{c20}). 
Eq.~(\ref{AA4}) does not contain the second order derivative of $h_{00}$ and 
other terms including the second order derivative with respect to time can be eliminated 
by the equations which we have already obtained. 
Therefore we obtain one more constraint and the system has five degrees of the freedom 
and there does not appear a ghost. 

Finally we investigate the relation between (\ref{d3}) and the non-minimal coupling in 
\cite{Buchbinder:1999ar}, which is given by 
\begin{align}
\frac{\xi}{4}Rh_{\alpha \beta} h^{\alpha \beta} + \frac{1-2\xi}{8}R h^2 
= \frac{R}{4} h_{\alpha \beta} h^{\alpha \beta}+\frac{R}{8} h^2 
- \frac{\xi -1}{4}R g^{\mu_1 \nu_1 \mu_2 \nu_2 } h_{\mu_1 \nu_1} h_{\mu_2 \nu_2} \, .
\label{d5}
\end{align}
The first two terms correspond to the shift of the mass term. 
By comparing (\ref{d5}) with the non-minimal coupling terms (\ref{rp1}) and (\ref{AA1}), 
we find that the expression (\ref{d5}) corresponds to the case that $\gamma =1$ in (\ref{AA1}). 
Then we find that in addition to the minimal coupling in \cite{Buchbinder:1999ar}, we can add 
the following non-minimal coupling, 
\begin{align}
\frac{\gamma}{2} C^{\mu \alpha \nu \beta}h_{\mu \nu} h_{\alpha \beta} \, .
\label{d10}
\end{align}
This term vanish on the (anti-)de Sitter space-time, which is conformally flat, but 
this term gives non-trivial contribution on the Schwarzchild (anti-)de Sitter spce-time, etc. 

\section{Derivative interaction}

In \cite{Hinterbichler:2013eza}, it has been shown that there are other kind of pseudo 
linear terms including derivative interaction although these terms do not correspond to 
any term in fully non-linear theory \cite{deRham:2013tfa}, 
\begin{align}
l\eta^{\mu_1 \nu_1 \mu_2 \nu_2 \mu_3 \nu_3 \mu_4 \nu_4} \partial_{\mu_1} \partial_{\nu_1} 
h_{\mu_2 \nu_2} \cdot h_{\mu_3 \nu_3 } h_{\mu_4 \nu_4}\, .
\label{AA5}
\end{align}
In this section, when we consider the coupling with gravity, we show that these terms always 
generate ghost by investigating if the derivative of $h_{00}$ with respect to time could appear in 
$\phi^{(2)\nu}$. 

If we include the derivative interaction terms, there appear the terms proportional to the 
curvature in the constraint  $\phi^{(2)\nu}$ due to the non-commutability of the 
covariant derivatives. 
In these terms, there appear the terms including the derivative of $h_{00}$ with respect to 
time and we need to cancel the terms by including the terms with non-minimal coupling to 
the action. Because there are not so many types of the non-minimal couplings, however, it 
is not trivial if we can cancel the terms including the derivative of $h_{00}$ with respect to 
time and in fact, we fail to cancel the terms. 

The derivative interaction term, 
\begin{align}
lg^{\mu_1 \nu_1 \mu_2 \nu_2 \mu_3 \nu_3 \mu_4 \nu_4} \nabla_{\mu_1} \nabla_{\nu_1} 
h_{\mu_2 \nu_2} \cdot h_{\mu_3 \nu_3 } h_{\mu_4 \nu_4}\, ,
\label{AA6}
\end{align}
give the following contribution to the equation, 
\begin{align}
E_D^{\mu \nu} \equiv 2l g^{(\mu \nu ) \mu_1 \nu_1 \mu_2 \nu_2 \mu_3 \nu_3 } 
\nabla_{\mu_1} \nabla_{\nu_1} h_{\mu_2 \nu_2} \cdot h_{\mu_3 \nu_3} 
+ l g^{(\mu \nu ) \mu_1 \nu_1 \mu_2 \nu_2 \mu_3 \nu_3 } \nabla_{\mu_1} h_{\mu_2 \nu_2} 
\cdot \nabla_{\nu_1} h_{\mu_3 \nu_3} \, .
\label{AA7}
\end{align}
In the expression of $\nabla_\nu E_D^{\mu \nu}$, the terms including the derivative of 
$h_{00}$ with respect to time are given by 
\begin{align}
\nabla_\nu E_D^{\mu \nu} \supset lg^{\mu \nu  \mu_1 \nu_1 \mu_2 \nu_2 \mu_3 \nu_3 } 
\left( -\frac12 \nabla_{\nu_2} h_{\mu_2 \sigma} + \nabla_\sigma h_{\mu_2 \nu_2} 
 -\nabla_{\mu_2} h_{\sigma \nu_2} \right) R_{\mu_1 ~ \nu \nu_1}^{~~\sigma} h_{\mu_3 \nu_3}
\, . \label{d8}
\end{align}
In the first term in the parentheses $(\ )$ in the r.h.s. of (\ref{d8}), we have used (\ref{d9b}). 
We now have following identities, 
\begin{align}
& g^{\mu \nu  \mu_1 \nu_1 \mu_2 \nu_2 \mu_3 \nu_3}C_{\mu_1 ~ \nu \nu_1}^{~~\sigma} 
= -6 (C^{\nu_2 \sigma (\mu \mu_2 }g^{\mu_3 ) \nu_3 } 
+ C^{\nu_3 \sigma (\mu \mu_3}g^{\mu_2 ) \nu_2} ) \, , \nn
& g^{\mu \nu  \mu_1 \nu_1 \mu_2 \nu_2 \mu_3 \nu_3} g_{\mu_1 \nu ~\nu_1}^{~~~~\sigma} 
=-2 g^{\mu \sigma \mu_2 \nu_2 \mu_3 \nu_3} \, , 
\label{d7}
\end{align}
In the first equation of (\ref{d7}), the parentheses $(\ )$ does not means the 
symmetrization but summing up by changing the indices in cyclic way, for example,  
\begin{align}
T_{(\alpha \beta \gamma )} \equiv \frac{1}{3} 
\left( T_{\alpha \beta \gamma} + T_{\beta \gamma \alpha} + T_{\gamma \alpha \beta} 
\right)\, .
\label{AA8}
\end{align}
By substituting (\ref{d7}) into (\ref{d8}) and using (\ref{d9}), we find that the terms 
proportional to $g^{\mu \sigma \mu_2 \nu_2 \mu_3 \nu_3}$ are pseudo-linear and therefore 
do not include the derivative of $h_{00}$ with respect to time. 
On the other hand, the terms including the Weyl tensor include the derivative of $h_{00}$ 
with respect to time and have the following forms, 
\begin{align}
\nabla_\nu E_D^{\mu \nu} \supset l \Bigl{\{} -C^{\mu \alpha 0 \beta} g^{00} 
+ C^{\alpha 0 \beta 0} g^{\mu 0} + C^{\mu 0 0 \alpha} g^{\beta 0} \Bigl{\}}
h_{\alpha \beta} \nabla_0 h_{00}\, .
\label{AA9}
\end{align}
Therefore in order to eliminate the ghost, we need to cancel the terms by including the 
terms with the non-minimal couplings but if we assume that the terms including the 
non-minimal couplings could have the following form, 
\begin{align}
c_1 C^{\mu \alpha \nu \beta} h_{\mu \nu} h_{\alpha \beta} h 
+ c_2 C^{\mu \alpha \nu \beta} h_{\mu \nu} h_\alpha^{~\lambda} h_{\lambda \beta} \, .
\label{d12}
\end{align}
Then the contribution to $\nabla_\nu E^{\mu \nu}$ from the term (\ref{d12}) are given by 
\begin{align}
\nabla_\mu E^{\mu \nu} \supset \left\{ (2c_1 + c_2 ) C^{\mu \alpha 0 \beta} g^{00} 
+ \left(2c_1 + c_2 \right) C^{0\alpha 0 \beta }g^{\mu 0} \right\} 
h_{\alpha \beta} \nabla_0 h_{00}
+ \left( \text{terms not including $\nabla_0 h_{00}$} \right) \, ,
\label{d11}
\end{align}
which tells that there cannot be cancellation. 
Therefore at least in the present formulation we cannot obtain the theory without ghost on 
the general Einstein manifold. 
We should note, however, that on the conformally flat space-time, where 
$C^{\mu \alpha \nu \beta} =0$, Eq.~(\ref{d8}) has the following form, 
\begin{align}
\nabla_\nu E_D^{\mu \nu} \supset -\frac{R}{12} 
g^{\mu \nu \mu_2 \nu_2 \mu_3 \nu_3}\nabla_\nu h_{\mu_2 \nu_2} h_{\mu_3 \nu_3}\, ,
\label{AA10}
\end{align}
and we obtain the ghost-free theory even if we do not include the non-minimal coupling 
because Eq.~(\ref{AA10}) does not include $\nabla_0 h_{00}$. 

We also have tried to eliminate the ghost by including the contribution from the non-minimal 
coupling in $\nabla_\mu \nabla_\nu E_D^{\mu \nu}$but we have not succeeded to construct 
$\phi^{(3)}$ on the general Einstein manifold. 

\section{Various non-minimal couplings}

We find that there are many kinds of minimal couplings which do not change the degrees of 
freedom and  do not generate a ghost. 
Because the scalar curvature is constant on the Einstein manifold and the terms without 
derivative do not violate the constraints, rather trivial terms are given by 
\begin{align}
R^m g^{\mu_1 \nu_1 \cdots \mu_n \nu_n} h_{\mu_1 \nu_1 } \cdots h_{ \mu_n \nu_n}\, .
\end{align}
In the previous sections, we also found the term proportional to the Weyl tensor in 
(\ref{d10}). 
By using the Weyl tensor, we can construct various non-minimal couplings which do not 
generate the ghost. 
For example, if we add the term in (\ref{d12}) to the action which includes only quadratic 
terms and also do not include ghost, as clear from (\ref{d11}), if we choose $c_2 = -2 c_1$, 
the non-minimal coupling (\ref{d12}) does not change the degrees of freedom and do not 
generate the ghost. 
When we neglect the over all factor, the terms of the non-minimal couplings are given by 
\begin{align}
C^{\mu \alpha \nu \beta}& h_{\mu \nu} h_{\alpha \beta} h 
- 2 C^{\mu \alpha \nu \beta} h_{\mu \nu} h_{\alpha \lambda} h^\lambda_{~ \beta} 
\nn 
&=\left( C^{\mu_1 \mu_2 \nu_1 \nu_2} g^{\mu_3 \nu_3} 
+ C^{\mu_1 \mu_2 \nu_2 \nu_3} g^{\mu_3 \nu_1}
+C^{\mu_1 \mu_2 \nu_3 \nu_1} g^{\mu_3 \nu_2} \right) 
h_{\mu_1 \nu_1} h_{\mu_2 \nu_2} h_{\mu_3 \nu_3} \nn
&=\frac{1}{2\cdot 3!} \delta^{\mu_1 ~~\mu_2 ~~ \mu_3}_{~~\rho_1 ~~\rho_2 ~~\rho_3 }  \delta^{\nu_1 ~~\nu_2 ~~ \nu_3}_{~~\sigma_1 ~~\sigma_2 ~~\sigma_3 } 
C^{\rho_1 \rho_2 \sigma_1 \sigma_2} g^{\rho_3 \sigma_3} h_{\mu_1 \nu_1} h_{\mu_2 \nu_2} 
h_{\mu_3 \nu_3} \, .
\label{d12b}
\end{align}
In (\ref{d12b}), under the exchange of the indices, the tensor 
$\delta^{\mu_1 ~~\mu_2 ~~ \mu_3}_{~~\rho_1 ~~\rho_2 ~~\rho_3 }  \delta^{\nu_1 ~~\nu_2 ~~ \nu_3}_{~~\sigma_1 ~~\sigma_2 ~~\sigma_3 } 
C^{\rho_1 \rho_2 \sigma_1 \sigma_2} g^{\rho_3 \sigma_3} $ has a structure of symmetry 
which is similar to that of $g^{\mu_1 \nu_1 \mu_2 \nu_2 \mu_3 \nu_3}$ in the pseudo 
linear theory as in (\ref{d10}). 
On the Einstein manifold, because $R_{\mu \nu}$ is proportional to $g_{\mu \nu}$, the 
terms including $R_{\mu \nu}$ is proportional to the tensor 
$g^{\mu_1 \nu_1 \mu_2 \nu_2 \mu_3 \nu_3}$ but because the Weyl tensor do not 
proportional to the tensor including $g_{\mu\nu}$, we can make a non-trivial tensor. 
If we only include the terms proportional the first power of the curvature in the non-minimal 
couplings, the possible tensor for the coefficients could be given by using one Weyl tensor 
$C^{\mu \alpha \nu \beta}$ and $n$ $g^{\mu \nu}$ as follows, 
\begin{align}
&\delta^{\mu_1  ~~ \mu_2 \cdots \mu_{n+2}}_{~~\rho1 ~~\rho_2 \cdots \rho_{n+2}}
\delta^{\nu_1  ~~ \nu_2 \cdots \nu_{n+2}}_{~~\sigma_1 ~~\sigma_2 \cdots \sigma_{n+2}} 
C^{\rho_1 \rho_2  \sigma_1 \sigma_2}
g^{\rho_{3}\sigma_{3}}  \cdots g^{ \rho_{n+2} \sigma_{n+2}} \notag \\
&\sim
\delta^{\mu_1  ~~ \mu_2 \cdots \mu_{n+2}}_{~~\rho1 ~~\rho_2 \cdots \rho_{n+2}}
\delta^{\nu_1  ~~ \nu_2 \cdots \nu_{n+2}}_{~~\sigma_1 ~~\sigma_2 \cdots \sigma_{n+2}} 
C^{\rho_1 \rho_2  \sigma_1 \sigma_2}
g^{\rho_{3}\sigma_{3}  \cdots \rho_{n+2} \sigma_{n+2}} \, .
\label{d13}
\end{align}
If we include the higher power of the curvature tensors, we may obtain more kinds of the 
tensors. 
In case of the derivative interaction terms, there could occur the difficulties similar to those 
in the last section even if we includes the Weyl tensor and possible terms could be given by 
contracting the indices by using $h_{\mu \nu}$. 
For the terms without the derivative interactions, it could be manifest that the terms do 
not generate a ghost. 
In four dimensions, the possible non-minimal couplings are given by the following three 
terms, 
\begin{align}
& C^{\mu_1 \mu_2 \nu_1 \nu_2} h_{\mu_1 \nu_1} h_{\mu_2 \nu_2 } \, , \nn
&\delta^{\mu_1 ~~\mu_2 ~~ \mu_3}_{~~\rho_1 ~~\rho_2 ~~\rho_3 } 
 \delta^{\nu_1 ~~\nu_2 ~~ \nu_3}_{~~\sigma_1 ~~\sigma_2 ~~\sigma_3 } 
C^{\rho_1 \rho_2 \sigma_1 \sigma_2} g^{\rho_3 \sigma_3} h_{\mu_1 \nu_1} h_{\mu_2 \nu_2} 
h_{\mu_3 \nu_3} \nn
& \delta^{\mu_1 ~~\mu_2 ~~ \mu_3 ~~\mu_4}_{~~\rho_1 ~~\rho_2 ~~\rho_3 ~~\rho_4} 
 \delta^{\nu_1 ~~\nu_2 ~~ \nu_3 ~~\nu_4}_{~~\sigma_1 ~~\sigma_2 ~~\sigma_3 ~~\sigma_4} 
C^{\rho_1 \rho_2 \sigma_1 \sigma_2} g^{\rho_3 \sigma_3 \rho_4 \sigma_4} 
h_{\mu_1 \nu_1} h_{\mu_2 \nu_2}  h_{\mu_3 \nu_3} h_{\mu_4 \nu_4} \, .
\end{align}

\section{Summary}

In this paper, we consider the model where a new massive spin two model 
\cite{Ohara:2014vua} couples with gravity. 
Although the model proposed in \cite{Ohara:2014vua} is ghost-free on the Minkowski 
space-time, the properties on the curved space-time is not so obvious.
In fact, Buchbinder et al. have shown that the Fierz-Pauli theory minimally coupled with 
gravity is not ghost-free \cite{Buchbinder:1999ar} and they have obtained a ghost-free 
theory in the background of the Einstein manifold by adding two non-minimal coupling 
terms. 
We also considered the model proposed in \cite{Ohara:2014vua} on the curved space-time 
by adding two non-minimal coupling terms as in the case of the Fierz-Pauli theory. 
Although the calculations become rather complicated and tedious, we have shown that 
the obtained model does not include ghost. 
Furthermore we investigated if the derivative interaction on 
curved space-time can be consistently formulated. 
Unfortunately, the derivative term induces a ghost even 
if we take the non-minimal coupling terms into account. Hence, the method of constructing 
a new spin two theory with the anti-symmetric tensor on Minkowski space-time cannot 
be extended to the theory on curved space-time by simply replacing $\eta^{\mu \nu}$ with $g^{\mu \nu}$.
On the other hand, this pseudo-linear approach leads to the discovery of new non-minimal coupling 
terms.

A motivation to consider this model on the curved background is applications to the 
cosmology and black hole physics.  
When we consider the cosmology, usually we assume the homogeneity and the isotropy 
of the spacial part of the universe. 
Furthermore in order to generate the accelerating expansion of the universe, we often consider 
the condensation of the field like inflaton. 
In case of the scalar field model, the condensation 
of the scalar field does not violate the isotropy although the condensation of the abelian 
vector field violates the isotropy. 
In fact, the condensation of the spacial components $A_i$ ($i=1,2,3$) in the vector field makes a 
special direction in the space. 
On the other hand, the condensation of the temporal component $A_0$ often 
conflicts with the gauge invariance because we can usually choose the gauge condition where the 
temporal component vanishes if there remains the gauge symmetry. 
In case of the non-abelian gauge theory, however, the condensation of the vector field does not 
conflict with the isotropy. 
Non-abelian gauge symmetry always include $SU(2)$ or $SO(3)$ gauge symmetry as a subgroup. 
Then we may consider the condensation of the vector field $A^a_i=A \delta^a_{\ i}$, where 
$a=1,2,3$ is the index of $SU(2)$ or $SO(3)$. 
The condensation breaks both of the rotational symmetry, which is $SO(3)$ symmetry and 
$SU(2)$ or $SO(3)$ gauge symmetry simultaneously because the condensation is not invariant 
under the rotation, ${A^a_i}' \equiv O_{Ri}^{\ \ \ j} A^a_j \neq A^a_i$, nor gauge transformation with 
a constant parameter, ${A^a_i}'  \equiv O_{G\ b}^{\ \ a} A^b_i \neq A^a_i$, where 
$O_{Rj}^{\ \ i}$ and $O_{G\ b}^{\ \ a}$ are elements of $SO(3)$, corresponding to the rotation 
and the gauge transformation with a constant parameter, respectively. 
We should note that the diagonal symmetry is preserved. 
In fact, the condensation of the vector field is invariant 
${A^a_i}' \equiv O_{Ri}^{\ \ \ j} O_{G\ b}^{\ \ a} A^b_j = A^a_j$ if we choose $O_R$ to be equal to 
the matrix of $O_G$, 
Then  we may identify this diagonal symmetry as a new rotational symmetry and the isotropy of 
of the spacial part is preserved. 
On the other hand, in case of the massive spin two field, which is the rank 2 symmetric tensor, 
the condensation of the trace part (or $(t,t)$ component, or the trace of the spacial part) 
does not violate the isotropy. 
Therefore we can consider easily the condensation of the rank 2 symmetric tensor in order to 
generate the expansion of the universe. 

In case of the massive gravity model, by using the condensation, cosmology has been 
investigated by considering the decoupling limit in \cite{deRham:2010tw} and there have 
been  many works about the cosmology in the massive gravity models 
\cite{Kluson:2012zz,Kluson:2012wf,Hassan:2011ea,D'Amico:2011jj} 
and in the bigravity models 
\cite{Damour:2002wu,Volkov:2011an,vonStrauss:2011mq,
Berg:2012kn,Nojiri:2012zu,Akrami:2012vf, Nojiri:2012re,Bamba:2013fha,AKMS-TSK,Solomon:2014dua,Konnig:2014xva}. 
It could be also interesting to investigate the black hole entropy as in 
\cite{Katsuragawa:2013bma,Katsuragawa:2013lfa}. 

\section*{Acknowledgments} 

The work is supported by the JSPS Grant-in-Aid for Scientific 
Research (S) \# 22224003 and (C) \# 23540296 (S.N.). 

\appendix

\section{Properties of $g^{\mu_1 \nu_1 \cdots \mu_n \nu_n}$}
In this appendix, we list the properties of $g^{\mu_1 \nu_1 \cdots \mu_n \nu_n}$. 
Note that the properties below are held on arbitrary space-time.
\subsection{Definition}
First, we define the tensor $g^{\mu_1 \nu_1 \cdots \mu_n \nu_n}$ as
\begin{align}
g^{\mu_1 \nu_1 \cdots \mu_n \nu_n} &\equiv g^{\mu_1 \nu_1} g^{\mu_2 \nu_2} g^{\mu_3 \nu_3} \cdots g^{\mu_n \nu_n} -
  g^{\mu_1 \nu_2} g^{\mu_2 \nu_1} g^{\mu_3 \nu_3} \cdots g^{\mu_n \nu_n}+ \cdots \nn
 &=\frac{-1}{(D-n)!} E^{\mu_1 \mu_2 \cdots \mu_n \sigma_{n+1} \cdots \sigma_D } E^{\nu_1 \nu_2 \cdots \nu_n}_{~~~~~~~~~~\sigma_{n+1} \cdots \sigma_D } \, . \label{Ap1}
\end{align}
Here $D$ denotes the dimension of the space-time and the totally anti-symmetric tensor $E^{\mu_1 \mu_2 \cdots \mu_n}$ is defined as   
\begin{align}
E^{\mu_1 \mu_2 \cdots \mu_D} \equiv \frac{1}{\sqrt{-g}} \epsilon^{\mu_1 \mu_2 \cdots \mu_D} \, .
\end{align}
with the totally anti-symmetric Levi-Civita tensor density
\[
\epsilon^{\mu_1 \mu_2 \cdots \mu_D}=\left\{
\begin{array}{l}
+1\;  \mbox{if} (\mu_1 \mu_2 \cdots \mu_D) \: \mbox{is an even permutation of} (0123\cdots) \\
 -1\; \mbox{if} (\mu_1 \mu_2 \cdots \mu_D) \: \mbox{is an even permutation of} (0123\cdots) \\
0 \; \mbox{otherwise}  
\end{array}
\right.
\]
In the following, we call the tensor $g^{\mu_1 \nu_1 \cdots \mu_n \nu_n}$ the pseudo-linear tensor.

Finally, we summarize the symmetric property of the pseudo-linear tensor.
\begin{align}
&\mu_i \longleftrightarrow \mu_j \text{ :anti-symmetric} \nn
&\nu_i \longleftrightarrow \nu_j \text{ :anti-symmetric} \nn
&(\mu_i , \nu_i ) \longleftrightarrow (\mu_j , \nu_j ) \text{ :symmetric} \nn
&\{ \mu_i \} \longleftrightarrow \{ \nu_i \} \text{ :symmetric}
\end{align}

\subsection{Useful relations}

The contraction of a pair of indices $\mu_n$ and $\nu_n$ leads to the following relation:
\begin{align}
\label{rep4}
{g^{\mu_1 \nu_1 \cdots \mu_{n-1} \nu_{n-1} \mu_n }}_{\mu_n}= (D-n+1) g^{\mu_1 \nu_1 \cdots \mu_{n-1} \nu_{n-1} } \, .
\end{align}

The pseudo linear tensor can be expanded in terms of the lower rank tensor:
\begin{align}
g^{\mu_1 \nu_1 \cdots \mu_n \nu_n} &= \delta^{\nu_1 ~\nu_2 \cdots \nu_n}_{~\lambda _1~ \lambda_2 \cdots \lambda_n} 
g^{\mu_1 \lambda_1} \cdots g^{\mu_n \lambda_n} \nn
&= \delta^{\nu_1 ~\nu_2 \cdots \nu_n}_{~\lambda _1~ \lambda_2 \cdots \lambda_n} \frac{1}{m! (n-m)!}
 g^{\mu_1 \lambda_1 \cdots \mu_m \lambda_m} g^{\mu_{m+1} \lambda_{m+1} \cdots \mu_n \lambda_n }  \, . \label{rep5}
\end{align}

For example, 
\begin{align}
g^{\mu_1 \nu_1 \mu_2 \nu_2 \mu_3 \nu_3} &= g^{\mu_1 \nu_1 } g^{\mu_2 \nu_2 \mu_3 \nu_3} + g^{\mu_1 \nu_2 } g^{\mu_2 \nu_3 \mu_3 \nu_1} 
+ g^{\mu_1 \nu_3} g^{\mu_2 \nu_1 \mu_3 \nu_2} \, , \nn
g^{\mu_1 \nu_1 \mu_2 \nu_2 \mu_3 \nu_3 \mu_4 \nu_4} &= g^{\mu_1 \nu_1 } g^{\mu_2 \nu_2 \mu_3 \nu_3 \mu_4 \nu_4} -
 g^{\mu_1 \nu_2 } g^{\mu_2 \nu_1 \mu_3 \nu_3 \mu_4 \nu_4} 
- g^{\mu_1 \nu_3} g^{\mu_2 \nu_2 \mu_3 \nu_1 \mu_4 \nu_4} - g^{\mu_1 \nu_4} g^{\mu_2 \nu_2 \mu_3 \nu_3 \mu_4 \nu_1} \, ,
\end{align}
(\ref{rep4}) and (\ref{rep5}) can be easily proven from (\ref{Ap1}).

We obtain other useful relations using the ADM variables $e_{ij}$, $N$, and $N_i$:
\begin{align}
& g^{0j i_1 j_1 i_2 j_2 \cdots i_n j_n } = \frac{1}{n!}
\delta^{j~j_1~j_2\cdots j_n}_{~k~k_1~k_2\cdots k_n} \frac{N^k}{N^2} 
e^{i_1 k_1 i_2 k_2 \cdots i_n k_n},\,  
\label{p1} \\
& g^{0j_1 i_1 0 i_2 j_2 i_3 j_3 \cdots i_n j_n } = \frac{1}{N^2} e^{i_1 j_1 i_2 j_2 \cdots i_n j_n} 
\, . \label{p2}
\end{align}
Here $e^{i_1 j_1 i_2 j_2 \cdots i_n j_n}$ is anti-symmetrization of the product 
$e^{i_1 j_1}e^{ i_2 j_2} \cdots e^{ i_n j_n}$ with respect to $j_i$.
\begin{align}
e^{i_1 j_1 i_2 j_2 \cdots i_n j_n} \equiv e^{i_1 j_1}e^{ i_2 j_2} \cdots 
e^{ i_n j_n} -e^{i_1 j_2}e^{ i_2 j_1} \cdots e^{ i_n j_n}+ \cdots \, .
\end{align}
Let us prove the identities (\ref{p2}). 
Just for convenience, we define the following tensors, 
\begin{align}
\tilde{g}^{\mu_1 \nu_1 \cdots \mu_n \nu_n} &\equiv \frac{1}{n!} 
g^{\mu_1 \nu_1 \cdots \mu_n \nu_n} 
=\tilde{\delta}^{\nu_1 \cdots \nu_n }_{~\lambda_1 \cdots \lambda_n} g^{\mu_1 \lambda_1} 
\cdots g^{\mu_n \lambda_n} \nn
& = \frac{1}{n!} (g^{\mu_1 \nu_1} g^{\mu_2 \nu_2} g^{\mu_3 \nu_3} \cdots g^{\mu_n \nu_n}
 - g^{\mu_1 \nu_2} g^{\mu_2 \nu_1} g^{\mu_3 \nu_3} \cdots g^{\mu_n \nu_n} + \cdots )\, ,
\nn
\tilde{e}^{i_1 j_1 \cdots i_n j_n} &\equiv \frac{1}{n!} e^{i_1 j_1 \cdots i_n j_n}
=\tilde{\delta}^{j_1\cdots j_n}_{~k_1\cdots k_n} e^{i_1 k_1} \cdots e^{i_n k_n}\, .
\end{align}
Therefore, we can easily prove (\ref{p1}) as follows, 
\begin{align}
\tilde{g}^{0j i_1 j_1 i_2 j_2 \cdots i_n j_n } &
= \tilde{\delta}^{j~j_1\cdots j_n}_{~\lambda ~\lambda_1\cdots \lambda_n}
g^{0\lambda} g^{i_1 \lambda_1 } \cdots g^{i_n \lambda_n} 
= \tilde{\delta}^{j~j_1\cdots j_n}_{~k ~k_1\cdots k_n}
g^{0k} g^{i_1 k_1 } \cdots g^{i_n k_n} \nn
&=\tilde{\delta}^{j~j_1\cdots j_n}_{~k ~k_1\cdots k_n}
 \frac{N^k}{N^2} \left( e^{i_1 k_1 } -\frac{N^{i_1} N^{k_1}}{N^2} \right) \cdots 
\left( e^{i_n k_n } -\frac{N^{i_n} N^{k_n}}{N^2} \right) \nn
&=\tilde{\delta}^{j~j_1\cdots j_n}_{~k ~k_1\cdots k_n}
\frac{N^k}{N^2} e^{i_1 k_1 } \cdots e^{i_n k_n } 
=\tilde{\delta}^{j~j_1\cdots j_n}_{~k ~k_1\cdots k_n}
\frac{N^k}{N^2} \tilde{e}^{i_1 k_1 \cdots i_n k_n} \, .
\end{align}
Furthermore, we can prove (\ref{p2}) by using mathematical induction. 
\begin{enumerate}
\item $n=1$ case
\begin{align}
g^{0ji0}= \frac{N^i}{N^2} \frac{N^j}{N^2} - \frac{1}{N^2} \left( e^{ij} - \frac{N^i N^j}{N^2} 
\right) = \frac{e^{ij}}{N^2} \, .
\end{align}
\item $n=m$ case \\
If we assume,
\begin{align}
g^{0j_1 i_1 0 i_2 j_2 i_3 j_3 \cdots i_{m-1} j_{m-1} } 
= \frac{1}{N^2} e^{i_1 j_1 i_2 j_2 \cdots i_{m-1} j_{m-1}}\, ,
\label{p4}
\end{align}
then we find 
\begin{align}
\tilde{g}^{0j_1i_10i_2 j_2 \cdots i_m j_m}
=& \tilde{\delta}^{j_1~0~j_2\cdots j_m}_{~\lambda_1 ~\lambda ~\lambda_2 \cdots 
\lambda_m} 
g^{i_m \lambda_m} \tilde{g}^{0\lambda_1 i_1 \lambda  i_2 \lambda_2 \cdots i_{m-1} 
\lambda_{m-1} } \nn
=&\frac{1}{m+1} \left[ g^{i_m j_m} \tilde{g}^{0j_1 i_1 0  i_2  j_2 \cdots i_{m-1} j_{m-1} }
 -g^{i_m j_1}\tilde{g}^{0 j_m i_1 0  i_2  j_2 \cdots i_{m-1} j_{m-1} } \right. \nn
& -g^{i_m 0}\tilde{g}^{0 j_1 i_1 j_m  i_2  j_2 \cdots i_{m-1} j_{m-1} }
 -g^{i_m j_2}\tilde{g}^{0 j_1 i_1 0  i_2  j_m \cdots i_{m-1} j_{m-1} } \cdots \nn
& \left. -g^{i_m j_{m-1}}\tilde{g}^{0 j_n i_1 0  i_2  j_2 \cdots i_{m-1} j_m } \right] \nn
=& \frac{1}{m+1} \left[ m\tilde{\delta}^{j_1~j_2\cdots j_m}_{~k_1~k_2\cdots k_m} g^{i_m k_m} 
\tilde{g}^{0k_1 i_1 0  i_2  k_2 \cdots i_{m-1} k_{m-1} }
 -g^{i_m 0} \tilde{g}^{0j_1 i_1 j_m i_2 j_2 \cdots i_{m-1} j_{m-1}} \right] \nn
=& \frac{1}{m+1} \left[ \tilde{\delta}^{j_1~j_2\cdots j_m}_{~k_1~k_2\cdots k_m} g^{i_m k_m} 
\frac{1}{N^2} \tilde{e}^{i_1 k_1 i_2 k_2 \cdots i_{m-1} k_{m-1}}
 -g^{i_m 0} \tilde{g}^{0j_1 i_1 j_n i_2 j_2 \cdots i_{m-1} j_{m-1}}  \right] \nn
=& \frac{1}{m+1} \left[ \tilde{\delta}^{j_1~j_2\cdots j_m}_{~k_1~k_2\cdots k_m} e^{i_m k_m} 
\frac{1}{N^2} \tilde{e}^{i_1 k_1 i_2 k_2 \cdots i_{m-1} k_{m-1}} \right] \nn
=&\frac{1}{m+1}\frac{1}{N^2} \tilde{e}^{i_1 j_1 i_2 j_2 \cdots i_{m} j_{m}}\, .
\end{align}
We used the assumption (\ref{p4}) in the fourth line and also used equation (\ref{p1}) in 
the fifth line. 
\end{enumerate}
So we have proved equations (\ref{p1}) and (\ref{p2}).

\end{document}